\documentclass[conf]{new-aiaa}
\usepackage[utf8]{inputenc}

\usepackage{graphicx}
\usepackage{amsmath}
\usepackage[version=4]{mhchem}
\usepackage{siunitx}
\usepackage{longtable,tabularx}
\RequirePackage{subcaption}
\usepackage{pgfplots}
\usepackage{tikz}
\usepgfplotslibrary{external}
\DeclareUnicodeCharacter{2212}{−}
\usepgfplotslibrary{groupplots,dateplot}
\usetikzlibrary{patterns,shapes.arrows}
\pgfplotsset{compat=newest}
\usetikzlibrary{intersections}
\usepgfplotslibrary{fillbetween}
\pgfplotsset{compat=1.16}
\usepackage[utf8]{inputenc}

\usepackage{graphicx}
\usepackage{amsmath}
\usepackage[version=4]{mhchem}
\usepackage{siunitx}
\usepackage{currfile}
\usepackage{gincltex}
\usepackage{xstring}
\usepackage{subcaption}
\usepackage{tikz}
\usepackage{tikz-3dplot}
\usepackage{pgfplots}
\usepackage{keyval}

\DeclareUnicodeCharacter{2212}{−}
\usepgfplotslibrary{groupplots,dateplot}
\usetikzlibrary{patterns,shapes.arrows,arrows.meta}
\usetikzlibrary{patterns,shapes.arrows,angles,calc,quotes}
\pgfplotsset{compat=newest}
\usetikzlibrary{intersections}
\usepgfplotslibrary{fillbetween}
\pgfplotsset{compat=1.16}
\newcommand{\figPath}{.}

\makeatletter
\newlength{\lp@height}
\newlength{\lp@width}
\define@key{lineplot}{height}{\setlength \lp@height{#1}}
\define@key{lineplot}{width}{\setlength \lp@width{#1}}
\define@key{lineplot}{xmin}{\def \lp@xmin{#1}}
\define@key{lineplot}{xmax}{\def \lp@xmax{#1}}
\define@key{lineplot}{ymin}{\def \lp@ymin{#1}}
\define@key{lineplot}{ymax}{\def \lp@ymax{#1}}
\define@key{lineplot}{xlabel}{\def \lp@xlabel{#1}}
\define@key{lineplot}{ylabel}{\def \lp@ylabel{#1}}
\define@key{lineplot}{xtick}{\def \lp@xtick{#1}}
\define@key{lineplot}{ytick}{\def \lp@ytick{#1}}
\define@key{lineplot}{legend}{\def \lp@legend{#1}}
\setkeys{lineplot}{height=0.3\textwidth, %
 width=0.95\textwidth,
 xmin = 0,
 xmax = 100,
 ymax = 100,
 ymin = -10,
 xlabel =$x$,
 ylabel = $y$,
 xtick = \lp@xmax/5,
 ytick = \lp@ymax/5,
 legend = 0}

\newcommand{\lineplot}[2][]{
\begin{tikzpicture}
\setkeys{lineplot}{#1}
\ifnum \lp@legend=1
\begin{axis}[
 /pgf/number format/.cd,
 1000 sep={},
 height=\lp@height,
 width=\lp@width,
 legend cell align={left},
 legend style={font=\footnotesize, fill opacity=0.8, draw opacity=1, text opacity=1, draw=white!80!black},
 tick align=outside,
 tick pos=left,
 unbounded coords=jump,
 xlabel= \lp@xlabel,
 xmin=\lp@xmin, xmax=\lp@xmax,
 xtick style={color=black},
 xtick distance=\lp@xtick,
 ylabel=\lp@ylabel,
 ymin=\lp@ymin, ymax=\lp@ymax,
 ytick distance = \lp@ytick,
 ytick style={color=black}
 ]
 \input{#2}
 \end{axis}
\else
 \begin{axis}[
 /pgf/number format/.cd,
 1000 sep={},
 height=\lp@height,
 width=\lp@width,
 every axis legend/.code={\let\addlegendentry\relax},
 tick align=outside,
 tick pos=left,
 unbounded coords=jump,
 xlabel=\lp@xlabel,
 xmin=\lp@xmin, xmax=\lp@xmax,
 xtick style={color=black},
 xtick distance=\lp@xtick,
 ylabel=\lp@ylabel,
 ymin=\lp@ymin, ymax=\lp@ymax,
 ytick distance = \lp@ytick,
 ytick style={color=black}
 ]
 \input{#2}
 \end{axis}
\fi
\end{tikzpicture}
}
\makeatother

\makeatletter
\newlength{\lpr@height}
\newlength{\lpr@width}
\define@key{lineplotraw}{height}{\setlength \lpr@height{#1}}
\define@key{lineplotraw}{width}{\setlength \lpr@width{#1}}
\define@key{lineplotraw}{xmin}{\def \lpr@xmin{#1}}
\define@key{lineplotraw}{xmax}{\def \lpr@xmax{#1}}
\define@key{lineplotraw}{ymin}{\def \lpr@ymin{#1}}
\define@key{lineplotraw}{ymax}{\def \lpr@ymax{#1}}
\define@key{lineplotraw}{xlabel}{\def \lpr@xlabel{#1}}
\define@key{lineplotraw}{ylabel}{\def \lpr@ylabel{#1}}
\define@key{lineplotraw}{xtick}{\def \lpr@xtick{#1}}
\define@key{lineplotraw}{ytick}{\def \lpr@ytick{#1}}
\setkeys{lineplotraw}{height=0.3\textwidth, %
 width=0.95\textwidth,
 xmin = 0,
 xmax = 100,
 ymax = 100,
 ymin = -10,
 xlabel =$x$,
 ylabel = $y$,
 xtick = \lpr@xmax/5,
 ytick = \lpr@ymax/5}

\newcommand{\lineplotraw}[2][]{
\setkeys{lineplotraw}{#1}
\begin{axis}[
/pgf/number format/.cd,
 1000 sep={},
height=\lpr@height,
width=\lpr@width,
every axis legend/.code={\let\addlegendentry\relax},
tick align=outside,
tick pos=left,
unbounded coords=jump,
xlabel=\lpr@xlabel,
xmin=\lpr@xmin, xmax=\lpr@xmax,
xtick style={color=black},
xtick distance=\lpr@xtick,
ylabel=\lpr@ylabel,
ymin=\lpr@ymin, ymax=\lpr@ymax,
ytick distance = \lpr@ytick,
ytick style={color=black}
]
 \input{#2}}
\makeatother

\makeatletter
\newlength{\cp@height}
\newlength{\cp@width}

\define@key{colorplot}{height}{\setlength \cp@height{#1}}
\define@key{colorplot}{width}{\setlength \cp@width{#1}}
\define@key{colorplot}{xmin}{\def \cp@xmin{#1}}
\define@key{colorplot}{xmax}{\def \cp@xmax{#1}}
\define@key{colorplot}{ymin}{\def \cp@ymin{#1}}
\define@key{colorplot}{ymax}{\def \cp@ymax{#1}}
\define@key{colorplot}{cmin}{\def \cp@cmin{#1}}
\define@key{colorplot}{cmax}{\def \cp@cmax{#1}}
\define@key{colorplot}{xlabel}{\def \cp@xlabel{#1}}
\define@key{colorplot}{ylabel}{\def \cp@ylabel{#1}}
\define@key{colorplot}{xtick}{\def \cp@xtick{#1}}
\define@key{colorplot}{ytick}{\def \cp@ytick{#1}}
\define@key{colorplot}{figPath}{\def \cp@figPath{#1}}
\define@key{colorplot}{colorMap}{\def \cp@colorMap{#1}}
\define@key{colorplot}{colorbar}{\def \cp@colorbar{#1}}
\define@key{colorplot}{colorbarPos}{\def \cp@colorbarPos{#1}}
\define@key{colorplot}{clegend}{\def \cp@clegend{#1}}
\define@key{colorplot}{wt}{\def \cp@wt{#1}}
\define@key{colorplot}{wtHeight}{\def \cp@wtHeight{#1}}
\define@key{colorplot}{wtPos}{\def \cp@wtPos{#1}}
\define@key{colorplot}{wtTop}{\def \cp@wtTop{#1}}
\setkeys{colorplot}{height=0.3\textwidth, %
 width=0.95\textwidth,
 xmin = 0,
 xmax = 100,
 ymax = 200,
 ymin = 0,
 xlabel =$x$,
 ylabel = $y$,
 xtick = \cp@xmax/5,
 ytick = \cp@ymax/5,
 figPath = \figPath,
 colorMap = jet,
 colorbar=1,
 colorbarPos=horizontal,
 cmin=0,
 cmax=10,
 clegend=color,
 wt = 1,
 wtPos = 0,
 wtHeight = 100,
 wtTop = 0,
}

\newcommand{\colorplot}[2][]{
\setkeys{colorplot}{#1}
\def\@horizontal{horizontal}
\def\@vertical{vertical}
\def\figpath{\cp@figPath}
\begin{tikzpicture}
 \ifnum \cp@colorbar=1
 \ifx\cp@colorbarPos\@horizontal
 \begin{axis}[%
 colorbar=true,
 colormap/\cp@colorMap,
 colorbar \cp@colorbarPos,
 colorbar style ={%
 ylabel =\cp@clegend,
 ylabel style={rotate=-90}
 },
 /pgf/number format/.cd,
 1000 sep={},
 axis background/.style={fill=black},
 height=\cp@height,
 width=\cp@width,
 tick align=outside,
 tick pos=left,
 x grid style={white!69.0196078431373!black},
 xmin=\cp@xmin, xmax=\cp@xmax,
 xtick style={color=black},
 xtick distance= \cp@xtick,
 xlabel=\cp@xlabel,
 ylabel =\cp@ylabel,
 y grid style={white!69.0196078431373!black},
 ymin=\cp@ymin, ymax=\cp@ymax,
 ytick distance = \cp@ytick,
 ytick style={color=black},
 point meta min=\cp@cmin,
 point meta max=\cp@cmax,
 ]
 \input{#2}
 \else
 \begin{axis}[%
 colorbar=true,
 colormap/\cp@colorMap,
 colorbar \cp@colorbarPos,
 colorbar style ={%
 title=\cp@clegend,
 },
 /pgf/number format/.cd,
 1000 sep={},
 axis background/.style={fill=black},
 height=\cp@height,
 width=\cp@width,
 tick align=outside,
 tick pos=left,
 x grid style={white!69.0196078431373!black},
 xmin=\cp@xmin, xmax=\cp@xmax,
 xtick style={color=black},
 xtick distance= \cp@xtick,
 xlabel=\cp@xlabel,
 ylabel =\cp@ylabel,
 y grid style={white!69.0196078431373!black},
 ymin=\cp@ymin, ymax=\cp@ymax,
 ytick distance = \cp@ytick,
 ytick style={color=black},
 point meta min=\cp@cmin,
 point meta max=\cp@cmax,
 ]
 \input{#2}
 \fi
\else
\begin{axis}[%
/pgf/number format/.cd,
1000 sep={},
axis background/.style={fill=black},
height=\cp@height,
width=\cp@width,
tick align=outside,
tick pos=left,
x grid style={white!69.0196078431373!black},
xmin=\cp@xmin, xmax=\cp@xmax,
xtick style={color=black},
xtick distance= \cp@xtick,
xlabel=\cp@xlabel,
ylabel =\cp@ylabel,
y grid style={white!69.0196078431373!black},
ymin=\cp@ymin, ymax=\cp@ymax,
ytick distance = \cp@ytick,
ytick style={color=black},
point meta min=\cp@cmin,
point meta max=\cp@cmax,
]
\input{#2}
\fi

\ifnum \cp@wt=1
 \ifnum \cp@wtTop=0
 \def\pos{\cp@wtPos}
 \def\height{\cp@wtHeight}
 \addplot [thin] coordinates {(\pos +30, \cp@wtHeight-100) (\pos+30, \cp@wtHeight)};
 \addplot [thin] coordinates {(\pos, \height-50) (\pos, \height+50)};
 \addplot [thin] coordinates {(0, \height) (\pos+30, \height)};
 \else
 \def\pos{\cp@wtPos}
 \def\height{\cp@wtHeight}

 \addplot [thin] coordinates {(\pos, 0) (\pos+30,0 )};
 \addplot [thin] coordinates {(\pos, -50) (\pos,+50)};

 \fi
 \fi

\end{axis}
\end{tikzpicture}
}
\makeatother

\makeatletter
\newlength{\plot@height}
\newlength{\plot@width}

\define@key{plot}{height}{\setlength \plot@height{#1}}
\define@key{plot}{width}{\setlength \plot@width{#1}}
\define@key{plot}{xmin}{\def \plot@xmin{#1}}
\define@key{plot}{xmax}{\def \plot@xmax{#1}}
\define@key{plot}{ymin}{\def \plot@ymin{#1}}
\define@key{plot}{ymax}{\def \plot@ymax{#1}}
\define@key{plot}{xlabel}{\def \plot@xlabel{#1}}
\define@key{plot}{ylabel}{\def \plot@ylabel{#1}}
\define@key{plot}{xtick}{\def \plot@xtick{#1}}
\define@key{plot}{ytick}{\def \plot@ytick{#1}}
\define@key{plot}{figPath}{\def \plot@figPath{#1}}
\define@key{plot}{colorbar}{\def \plot@colorbar{#1}}
\define@key{plot}{clegend}{\def \plot@clegend{#1}}
\define@key{plot}{wt}{\def \plot@wt{#1}}
\define@key{plot}{wtHeight}{\def \plot@wtHeight{#1}}
\define@key{plot}{wtPos}{\def \plot@wtPos{#1}}
\define@key{plot}{wtTop}{\def \plot@wtTop{#1}}
\setkeys{plot}{height=0.3\textwidth, %
 width=0.95\textwidth,
 xmin = 0,
 xmax = 100,
 ymax = 200,
 ymin = 0,
 xlabel =$x$,
 ylabel = $y$,
 xtick = \plot@xmax/5,
 ytick = \plot@ymax/5,
 figPath = \figPath,
 colorbar=false,
 clegend=color,
 wt = 0,
 wtPos = 0,
 wtHeight = 100,
 wtTop = 0,
}

\newcommand{\plot}[2][]{
\setkeys{plot}{#1}
\def\figpath{\plot@figPath}
\begin{tikzpicture}
\def\colorb{\plot@colorbar}
\def\xmin{\plot@xmin}
\def\height{\plot@height}
\def\width{\plot@width}
\def\xmin{\plot@xmin}
\def\xmax{\plot@xmax}
\def\ymax{\plot@ymax}
\def\ymin{\plot@ymin}
\def\xlabel{\plot@xlabel}
\def\ylabel{\plot@ylabel}
\def\xtick{\plot@xtick}
\def\ytick{\plot@ytick}
\def\clegend{\plot@clegend}
\pgfkeys{/pgf/number format/.cd,%additional code
 use comma,%additional code
 1000 sep={},%additional code
}
\input{#2}
 \begin{axis}[stack plots=y,
 area style,
 enlarge x limits=false,
 width=\width,
 height=\height,
 xmin=\xmin,
 xmax=\xmax,
 ymin=\ymin,
 ymax=\ymax,
 hide axis,
 ]
 \ifnum \plot@wt=1
 \ifnum \plot@wtTop=0
 \def\position{\plot@wtPos}
 \draw (\position +30, \plot@wtHeight-100) -- (\position+30, \plot@wtHeight);
 \draw (\position, \plot@wtHeight-50) -- (\position, \plot@wtHeight+50);
 \draw (\position, \plot@wtHeight) -- (\position+30, \plot@wtHeight);
 \else
 \def\pos{\plot@wtPos}
 \addplot [thin] coordinates {(\pos, 0) (\pos+30,0 )};
 \addplot [thin] coordinates {(\pos, -50) (\pos,+50)};
 \fi
 \fi
 \end{axis}
\end{tikzpicture}
}

\define@key{wt}{height}{\def \wt@height{#1}}
\define@key{wt}{pos}{\def \wt@pos{#1}}
\define@key{wt}{top}{\def \wt@top{#1}}
\setkeys{wt}{
 pos = 0,
 height = 100,
 top = 0,
}
\newcommand{\wt}[1][]{
\setkeys{wt}{#1}
 \def\pos{\wt@pos}
\ifnum \wt@top=0
 \draw[thick] (\pos +30, \wt@height-100) -- (\pos+30, \wt@height);
 \draw[thick] (\pos, \wt@height-50) -- (\pos, \wt@height+50);
 \draw[thick] (\pos, \wt@height) -- (\pos+30, \wt@height);
\else
 \addplot [thin] coordinates {(\pos, 0) (\pos+30,0 )};
 \addplot [thin] coordinates {(\pos, -50) (\pos,+50)};
\fi
}
\makeatother

\title{
\begin{flushleft}
\small{
\normalfont{
Jules Colas, Ariane Emmanuelli, Didier Dragna, Richard Stevens and Philippe Blanc-Benon,
Effect of a 2D Hill on the Propagation of Wind Turbine Noise, 
AIAA 2022, 2923 (2022).\\
See {\color{blue}https://doi.org/10.2514/6.2022-2923 }\\ 
Published by the American Institute of Aeronautics and Astronautics, Inc}}\\ \vspace{0.5cm}
\end{flushleft} 
Effect of a 2D Hill on the Propagation of Wind Turbine Noise}

\author{J. Colas, A. Emmanuelli, D. Dragna}
\affil{Univ Lyon, École Centrale de Lyon, INSA Lyon, Université Claude Bernard Lyon I, CNRS, Laboratoire de Mécanique des Fluides et d’Acoustique, UMR 5509, 36 Avenue Guy de Collongue, F-69134, Écully, France}
\author{R.J.A.M. Stevens}
\affil{Physics of Fluids Group, Max Planck Center Twente for Complex Fluid Dynamics, J. M. Burgers Center for Fluid Dynamics and MESA+ Research Institute, University of Twente, P. O. Box 217, 7500 AE Enschede, The Netherlands}
\author{P. Blanc-Benon}
\affil{Univ Lyon, École Centrale de Lyon, INSA Lyon, Université Claude Bernard Lyon I, CNRS, Laboratoire de Mécanique des Fluides et d’Acoustique, UMR 5509, 36 Avenue Guy de Collongue, F-69134, Écully, France}

\definecolor{blue}{rgb}{0.277, 0.607, 0.808}
\definecolor{UBCblue}{rgb}{0.277, 0.607, 0.808}

\begin{document}

\maketitle

\begin{abstract}
 A good understanding of wind turbine noise propagation is relevant to better measure the impact of turbines on the environment.
 In this study, we developed numerical simulations to study the impact of a 2D hill on the sound propagation for a turbine in non flat terrain.
 The simulations employed a propagation model derived from the linearized Euler equations, solved in a moving frame following the wavefront.
 The wind turbine noise was modeled using a point source approach.
 The flow structure in the atmospheric boundary layer and the wind turbine wake were obtained from large-eddy simulations.
 We find that the wind turbine wake acts as a waveguide that focuses sound propagation towards the ground.
 This effect is most pronounced for the flat terrain case and when the turbine is placed on the hill.
 The sound attenuation is more localized and closer to the turbine when the wind turbine is placed on the hilltop than in the flat case.
 As the wind speed increases, the sound focusing is observed closer to the turbine.
 For a turbine placed in front of the hill, the sound propagation is mainly determined by the flow over the hill.
 This demonstrates that terrain topography can have surprising effects on wind turbine noise propagation.
\end{abstract}

\section*{Introduction}

The increase in demand for renewable energy and the low power density characteristic of wind turbines lead to the development of extended wind farms.
One of the main obstacles to achieve such installations is public acceptance due to noise annoyance issues.
Hence, accurate models for noise prediction from single wind turbines and wind farms are necessary to extend the use of wind energy and achieve sustainable energy production worldwide.
In recent years, significant efforts have been made in order to characterize wind turbine noise.
This includes the study of the wind turbine inflow, the phenomena of noise emission, and the effect of the propagation medium on the far field noise.

Noise produced by wind turbines can propagate over several kilometers.
Aeroacoustic noise, which is caused by interaction of the incoming wind with the blades of the turbines, is dominant.
The main origins of aeroacoustic noise are turbulent inflow noise and trailing edge noise.
This results in a broadband noise with amplitude modulation due to the movement of the blades, which is considered to be one of the main annoyance issues.
The noise production is greatly influenced by the geometry of the blades, and by the incoming wind profile and turbulence characteristics.
Several models have been developed in order to simulate wind turbine noise production.
Point source approaches are commonly used by \citet{lee_prediction_2016} and \citet{prospathopoulos_application_2007}. However more complex approaches, taking into account an extended aeroacoustic sound source, were also developed \cite{cotte_extended_2019}.

\begin{figure}[h]
\centering
 \includegraphics[width = 0.8\textwidth]{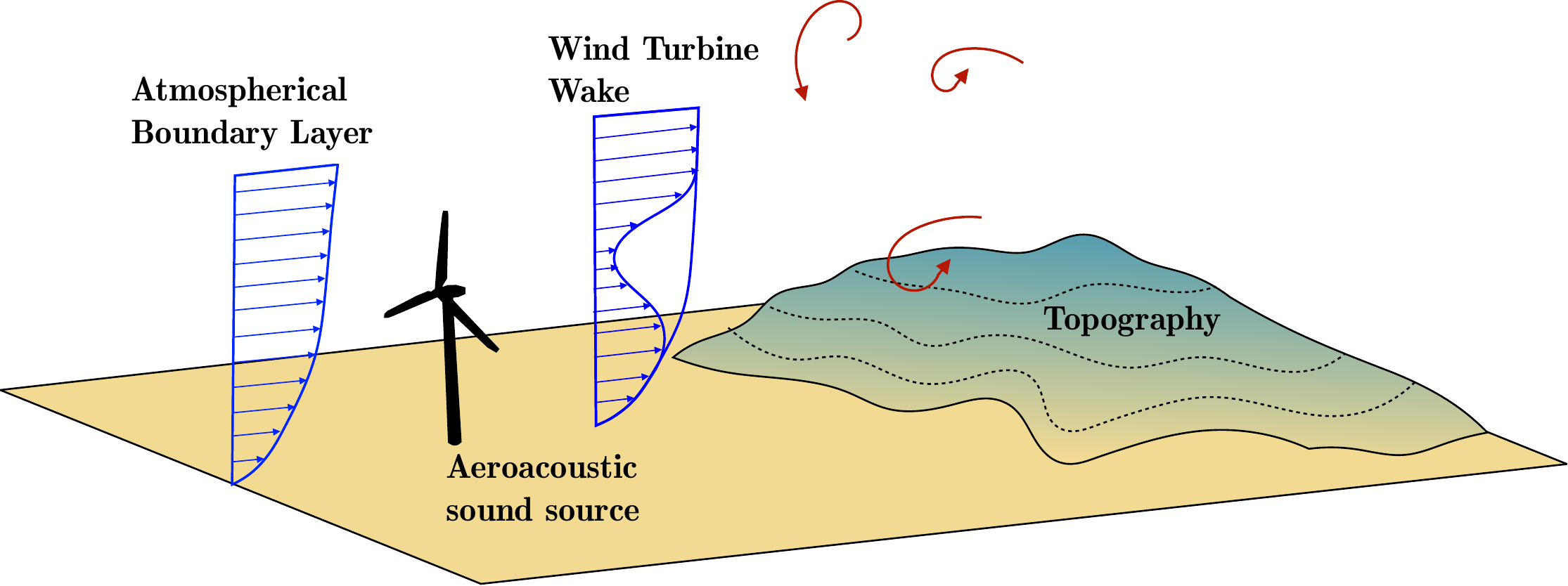}
\vspace{-0.2cm} \caption{Sketch of the of the general wind turbine noise propagation question}
 \label{fig:context}
\end{figure}

Atmospheric conditions greatly influence wind turbine noise perceived at long distances.
Several numerical methods have been developed in order to study these phenomena.
They are generally based on propagating sound through an already generated flow fields, either analytically or numerically.
The effect of wind velocity and temperature profiles was studied by \citet{barlas_variability_2018} and \citet{heimann_3d-simulation_2018} as they influence noise generation mechanisms and noise propagation.
An important finding is that the wind turbine wake can act as a waveguide and creates a focusing zone near the ground.
Its position and intensity notably depends on the velocity deficit of the wind turbine wakes.
One of the main results \cite{barlas_effects_2017,heimann_3d-simulation_2018} is that far field amplitude modulation is greatly influenced by wind shear and turbulent intensity and that focalization zones (their existence, intensity and position) strongly depend on small variations in the meteorological conditions.

In addition to atmospheric effects, topography and ground effects are known to be key parameters in long range sound propagation.
Configurations where the wind turbine is positioned on top of a hill are quite common as it can improve the performance.
\citet{berg_bolund_2011} conducted experimental studies for a single wind turbine over a hill.
The performance of a wind turbine around a hill was also studied using numerical simulations \cite{liu_effects_2020}.
Hence, in these configurations, it becomes necessary to take into account topographical effects on noise production and propagation.
The influence of a 3D hill on both the inflow of the wind turbine and on the noise propagation was studied by \citet{heimann_sound_2018}.
In their work, the atmospheric flow perturbed by the hill and the wind turbine was computed using Large Eddy Simulation (LES).
The results were then used as an input for propagation simulations using a 3D ray-based sound particles model.
The results show that positioning the wind turbine on top of the hill can lead to the reduction of the overall sound pressure level (SPL) in the far field.
This can compensate the effect of the wake at ground level, which tends to enhance the overall SPL in the far field.
\citet{shen_advanced_2019} studied topography effects for a realistic case considering a wind farm over complex ground.
The noise levels were compared with a flat ground test case and with a simple logarithmic profile.
Both the topography and the turbulent flow characteristics were shown to be key components in predicting noise propagation.

This work aims at further investigating the impact of topography on wind turbine noise propagation.
The goal is to study the case of a 2D hill, i.e. a ridge, and several positions of a single wind turbine in order to understand the effect of topography and of the wake on sound propagation.
The position of the wind turbine with respect to the hill is expected to have an impact both on the flow around the wind turbine, especially the development of the wake, and on sound propagation, through reflection, refraction and scattering of acoustic waves.
In order to study these phenomena, a propagation method based on the linearized Euler equations (LEE) is developed.
The effect of the hill and of the wind turbine on the flow are taken into account through the mean flow values obtained from LES performed in the work of \citet{liu_effects_2020}.
To the authors' knowledge, the LEE has never been used for the long range propagation of wind turbine noise, essentially because it is very computationally demanding.
It is interesting in this case as more classical methods based on geometrical acoustics or parabolic equations could be limited in presence of topography.

The paper is organized as follows. In Sec. \ref{Methodo} the methodology is described including the derivation of the propagation model, the numerical scheme and the post processing. Then the cases studied and the numerical simulation performed are presented in Sec \ref{cases}.
The sound pressure levels due to the wind turbine are presented and discussed in Sec. IV, depending on the presence of the hill, the location of the wind turbine and the wind velocity at the hub height. Finally, concluding remarks are given in Sec. V.

\section{Methodology} \label{Methodo}

To consider meteorological and ground effects (Fig. \ref{fig:context}), wind turbine noise is predicted using numerical simulations. The propagation model is first detailed before presenting the source model.

\subsection{Propagation Model}
The propagation model is based on the numerical resolution of the linearized Euler equations (LEE). They are solved in a 2D curvilinear mesh to account for topography. Low-dispersive high-order finite-difference schemes are used for an accurate prediction at long range.
\begin{figure}
 \centering
 \includegraphics[width=\textwidth]{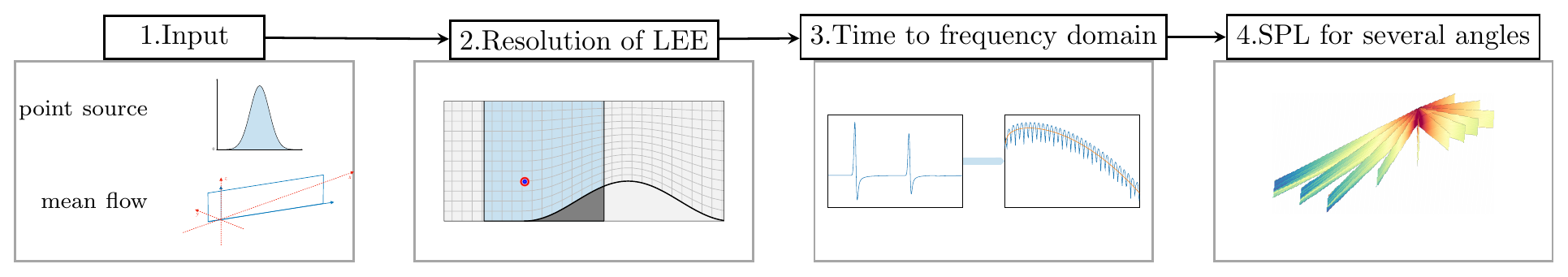}
\vspace{-0.8cm} \caption{Diagram of the complete methodology, see details in text.}
 \label{fig:method}
\end{figure}
A set of two equations is derived from the linearized Euler equations (LEE) for atmospheric acoustics without source terms \citep{ostashev_equations_2005}:

\begin{equation}
\begin{aligned}
&\frac{d p}{d t}+\rho_{0} c_0^{2} \nabla \cdot \mathbf{v} = 0 \\
&\frac{d \mathbf{v}}{d t}+(\mathbf{v} . \nabla) \mathbf{V}_{\mathbf{0}}+\frac{\nabla p}{\rho_{0}}=0
\end{aligned}
\end{equation}
with $p$ and $\mathbf{v}$ the acoustic pressure and velocity, $\rho_0$, $\mathbf{V_0}$ and $c_0$ the mean density, velocity and sound speed. A conservative formulation of this system of equations is then derived for the transformation of the coordinate system from Cartesian $(\mathbf{e_x},\mathbf{e_z})$ to curvilinear $(\mathbf{e_{\xi}},\mathbf{e_{\eta}})$ :

\begin{equation}
\label{eq-conservative}
\frac{\partial}{\partial t}\left(\frac{\mathbf{U}}{J}\right)+\frac{\partial}{\partial \xi}\left(\frac{\xi_{x} \mathbf{E}+\xi_{z} \mathbf{F}}{J}\right)+\frac{\partial}{\partial \eta}\left(\frac{\eta_{x} \mathbf{E}+\eta_{z} \mathbf{F}}{J}\right) + \frac{\mathbf{H}}{J} =0
\end{equation}
where $\mathbf{U} = (p,\rho_0 v_x, \rho_0 v_z)^T$ is the vector of unknown, $J = (\xi_x \eta_z - \xi_z \eta_x)$ is the Jacobian of the transformation and $\mathbf{E}$, $\mathbf{F}$, $\mathbf{H}$ are the Eulerian fluxes defined by :

\begin{equation}
\label{eq-flux}
\mathbf{E}=\left(\begin{array}{c}
V_{0 x} p+\rho_{0} c_0^{2} v_{x} \\
V_{0 x} \rho_{0} v_{x}+p \\
V_{0 x} \rho_{0} v_{z}
\end{array}\right),\mathbf{F}=\left(\begin{array}{c}
V_{0 z} p+\rho_{0} c_0^{2} v_{z} \\
V_{0 z} \rho_{0} v_{x} \\
V_{0 z} \rho_{0} v_{z}+p
\end{array}\right), \mathbf{H}=\left(\begin{array}{c}
-p\left(\nabla \cdot \mathbf{V}_{\mathbf{0}}\right) \\
\rho_{0}(v . \nabla) V_{0 x} \\
\rho_{0}(v . \nabla) V_{0 z}
\end{array}\right)
\end{equation}
The curvilinear transformation proposed by \citet{gal-chen_use_1975} is used in order to follow the terrain elevation $h(\xi)$ :
\begin{equation}
\begin{aligned}
&x(\xi, \eta)=\xi \\
&z(\xi, \eta)=h(\xi)+\frac{\eta}{z_{\rm max }}\left[z_{\rm max }-h(\xi)\right]
\end{aligned}
\end{equation}
with $z_{\rm max}$ the height of the domain.
This leads to the simplification of Eq. (\ref{eq-conservative}) as $\xi_x = 1$ and $\xi_z = 0$.

 The mean flow quantities in Eq. (\ref{eq-flux}) can be set using either analytical profiles for wind, temperature and density or, in our case, previously computed aerodynamic simulations.
 The acoustic fluctuations are then computed inside the domain using a high-order, low-dispersive optimized numerical schemes (Fig. \ref{fig:method}.2).
 Time integration is performed using a fourth order six-step Runge-Kutta algorithm \cite{berland_low-dissipation_2006}, the spatial derivatives are discretized using a fourth order finite difference scheme over an 11 points stencil \cite{bogey_family_2004}.
 Finally, selective filters are used in order to remove grid to grid oscillations \citep{bogey_shock-capturing_2009}.
 This is necessary as grid to grid oscillations are not solved in a central finite difference schemes and could lead to numerical instabilities if not removed.

To simulate a more realistic ground, a broadband impedance condition \cite{troian_broadband_2017} is implemented at the bottom of the domain.
The impedance condition is developed in the time domain and integrated along the Runge-Kutta time stepping scheme.
This avoids computing at each time step the convolution resulting from the translation of the frequency-domain impedance boundary condition.
A small variation from the proposed method is implemented.
In \citet{troian_broadband_2017} the method is developed for the admittance condition, which lead to some instabilities in our case.
This issue is solved by implementing the exact same methodology for the reflection coefficient.
In order to simulate unbounded propagation at the top of the domain, a non-reflective boundary condition using convolutional perfectly matched layer (CPML) type is implemented \cite{cosnefroy_simulation_2019,petropoulos_reflectionless_2000}.
The transformation inside the absorbing layer is defined as such:
\begin{equation}
 \frac{\partial}{\partial z} \longrightarrow \frac{1}{\kappa+\sigma/i \omega} \frac{\partial}{\partial z}
\end{equation}
and
\begin{equation}
\begin{aligned}
\sigma(z-z_{\rm PML}) &=\sigma_{0}((z-z_{\rm PML}) / L)^{m} \\
\kappa(z-z_{\rm PML}) &=1+\left(\kappa_{0}-1\right)((z-z_{\rm PML}) / L)^{m} \\
\end{aligned}
\end{equation}
with $z_{\rm PML}$ the coordinate at the start of the layer and $L$ the length of the layer. This condition is integrated along the Runge-Kutta scheme in the same way the impedance boundary condition was \cite{cosnefroy_simulation_2019}.

While the mean flow quantities, the mesh coordinates and the Jacobian are computed in the entire domain, the acoustic variables and the boundary conditions are only computed in a moving frame that follows the wavefront.
This allows to greatly reduce computational cost as the propagation distance is large (around 15000 wavelengths).
When the frame moves forward inside the domain, the acoustic variables are initialized at zero on the right boundary.
At the left boundary the sound speed decreases over a few grid points, such that the moving frame speed becomes greater than sound speed in this layer.
Hence the reflected waves at the border propagate slower than the frame moves and they cannot reenter the domain.
This method is described by \citet{cosnefroy_simulation_2019} and used by \citet{emmanuelli_characterization_2021} where the entire frame was moving at supersonic speed.

\subsection{Atmospheric Flow and Wind Turbine Source Model}
Two inputs are needed for the propagation model: a source term, and the mean quantities defined in the Eq.~(\ref{eq-flux}).
In this work the mean flow quantities are taken from the study by \citet{liu_effects_2020} on the effect of a 2D steep hill on the performance of a wind turbine.
The flow is simulated using LES and an immersed boundary method for a truly neutral atmospheric boundary layer (ABL).
The wind turbine is modeled using an actuator disk model (ADM).
The LES results are normalized by the friction velocity $u^*$ and by the diameter of the turbine.
In this work the diameter and height of the turbine are set to 100~m and the roughness height at the ground is 0.01, which is representative for flow over smooth terrain.
The friction velocity is then set $u^*=0.452 $m.s$^{-1}$ such that the wind velocity at hub height is 8m.s$^{-1}$, which a typical value for wind turbines application.
From the LES, the mean 3D quantities are interpolated on a 2D plane of propagation (Fig. \ref{fig:method}.1).
A spline interpolation is used as the LES mesh grid is much coarser than the acoustic simulation one (20 times larger).
In the presented method only the mean velocity is used for the acoustic simulation, but other quantities could be exploited in the future.
For the density and the sound speed, constant values were taken: $\rho_0 = 1.2~$kg.m$^{-3}$ and $c_0=343~$m.s$^{-1}$.

In order to compute the sound pressure level in the domain, a pulse, located at the wind turbine's hub, is propagated inside the moving frame. The source term is defined as an initial Gaussian condition (Fig. \ref{fig:method}.1):

\begin{equation}
\mathbf{U}(t=0)=(\exp(-A_0R^2), 0, 0)^T
\end{equation}
where $R$ is the distance to the source, $A_0 = \log(2)/(5\Delta x)^2 $ and $\Delta x$ the LEE grid spacing.

%%%% Plot
\begin{figure}[t]
 \begin{subfigure}[b]{0.49\textwidth}
 \centering
 \includegraphics[width=0.95\textwidth]{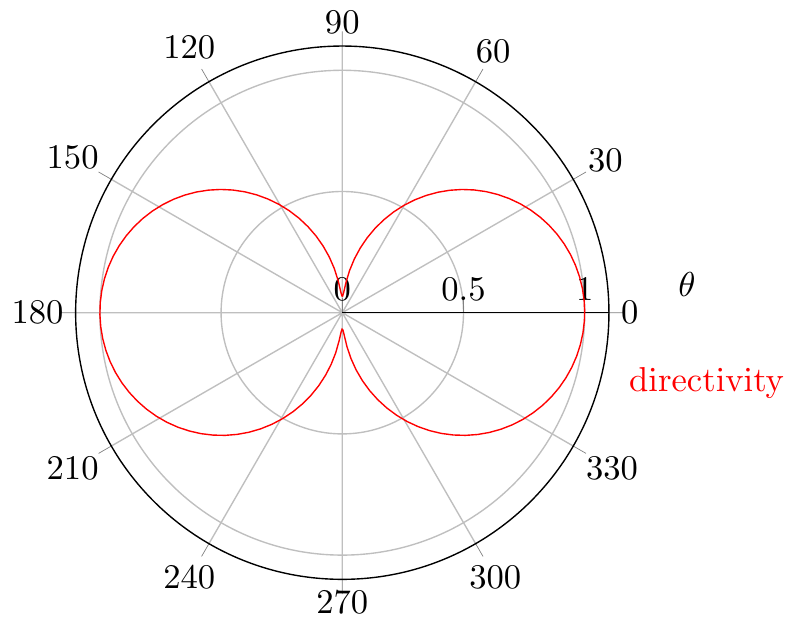}
% \vspace{0.6cm}
 \end{subfigure}
 \begin{subfigure}[b]{0.49\textwidth}
 \centering
 \includegraphics[width=0.95\textwidth]{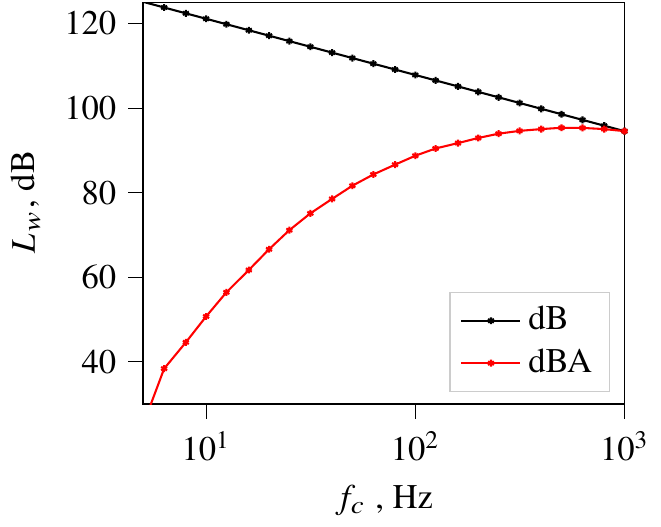}
% \vspace{0.6cm}
 \end{subfigure} 
% \begin{subfigure}[b]{0.49\textwidth}
 %\plot[xmin=5,
 %xmax=1000,%
 %ymin =30,
 %ymax=125,
 %width = 0.8\textwidth,
% height = 0.7\textwidth,
% xlabel = {$f_c$ , Hz},
% ylabel = {$L_w$, dB},
% xtick=10,
% ytick=20,
% colorbar=false,
% figPath=../figures_offline/flow/caseA,
% ]{../figures_offline/AIAA2/source/spectrum.tex}
% \end{subfigure}
\caption{ (left) Normalized horizontal directivity of the source and (right) sound power level of the source in dB and dBA.}
\label{fig:source}
\end{figure}
From the time domain solution, it is possible to recover a frequency domain solution (Fig. \ref{fig:method}.3). The sound pressure level relative to the free field can be derived from the time signal $p(t)$ recorded at one receiver. It is defined as:
\begin{equation}
 \Delta L (\omega,x,z) = 10 \log_{10}\left(\frac{|P(\omega,x,z)|^2}{|P_{f}(\omega,x,z)|^2}\right)
\end{equation}
with $P$ the Fourier transform of $p$ and $P_{f}$ the free field solution, i.e the solution for the same source term but without any mean flow or ground reflection. This quantity shows the effect of topography and of the mean flow on the propagation, without the influence of the source term, geometrical spreading or atmospheric absorption. Hence, once this quantity has been computed, the sound pressure level is recovered using \citep{salomons_computational_2001}:

\begin{equation}
 L_{p}(\omega,x,z)=L_{w}(\omega)-10 \log_{10} 4 \pi R^{2}-\alpha(\omega) R +\Delta L(\omega,x,z)
\end{equation}
with $L_w$ the sound power of the source, $R$ the distance between the receiver and the source and $\alpha$ the atmospheric absorption coefficient. For $L_w$ a point source model was implemented. A corrected dipole directivity in the $xy$ plane is taken from \citet{oerlemans_prediction_2009}, while the overall sound power empirical formulation proposed by \citet{moller_low-frequency_2011} is used, so that:
\begin{equation}
 L_{w}^{\rm tot}=11.0\log_{10} \left(\frac{W_{\rm nom}}{W_{\rm ref}}\right)+101.1
\end{equation}
with $W_{\rm nom}$ the nominal power of the wind turbine set to 2~MW and $W_{\rm ref}$ the reference power set to 1~MW. The sound power levels for each third octave band is computed in order to respect the overall sound power and a 4~dB per third octave band decrease \citep{kayser_estimation_2020}. This model gives the directivity and sound power per third octave band representative of a wind turbine (Fig. \ref{fig:source}).

Finally this methodology is applied for several propagation angles (Fig. \ref{fig:method}.4), and a map of the sound pressure level is obtained in the $xy$ plane around the wind turbine.
The numerical scheme is implemented in Fortran 90 with a parallelization in OpenMP.
The post processing and frequency analysis is performed using Python.
Simulations were run on the HPC cluster at the Ecole Centrale de Lyon.

\section{Case Presentation}
\label{cases}

\begin{figure}[b]
 \centering
 \includegraphics[ width=1.00\textwidth]{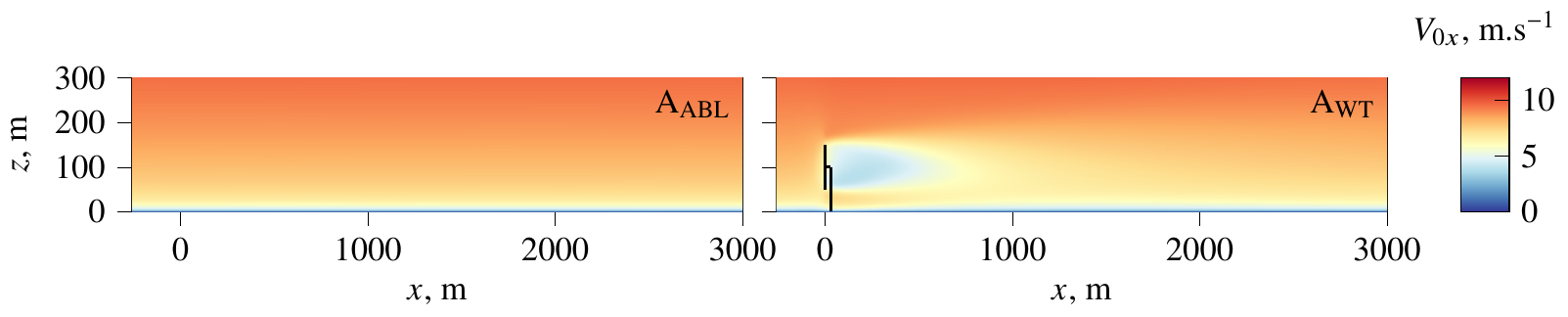}
\vspace{-0.8cm} \caption{Mean flow velocity field for the case A without (left) and with (right) wind turbine.}
\label{fig:caseA}
\end{figure}

\begin{figure}[tbp]
\centering
 \includegraphics[width=1.00\textwidth]{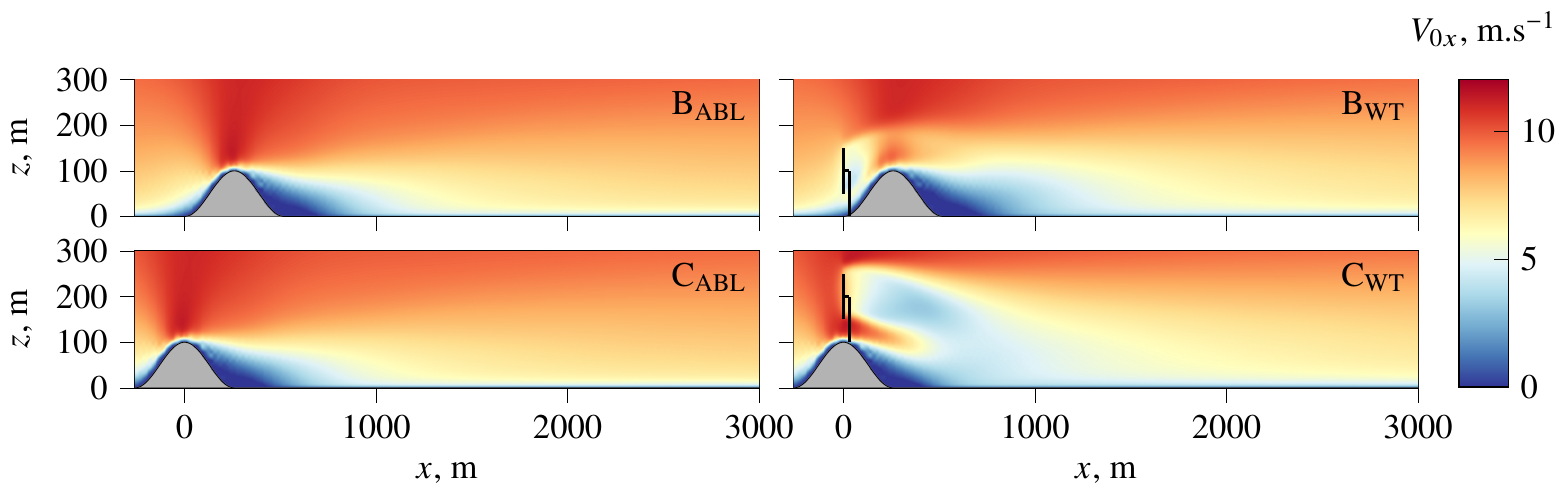}
\vspace{-0.8cm} \caption{Mean flow velocity field for the cases B and C with and without the wind turbine}
\label{fig:caseB-C}
\end{figure}

The work presented here focuses on three configurations that were previously computed by \citet{liu_effects_2020}.
For each configuration LES are run, for a truly neutral ABL, with and without the turbine inside the flow.
Hence the results are available for the mean flow perturbed by the wind turbine, as well as the unperturbed mean flow.
This cases are referred as $\rm X_{ABL}$ (without turbine) and $\rm X_{WT}$ (wind turbine).
This allows to investigate the effect of the wake on sound propagation with respect to the topography or the general mean flow.
The configurations are defined as such :
\begin{itemize}
 \item Case $\rm A_{ABL}$ and $\rm A_{WT}$: a baseline case of a wind turbine with a flat ground
 \item Case $\rm B_{ABL}$ and $\rm B_{WT}$: a wind turbine place in front of a 2D hill
 \item Case $\rm C_{ABL}$ and $\rm C_{WT}$: a wind turbine place on the top of a 2D hill
\end{itemize}
The same topography is used for configuration B and C. It is a 2D hill (constant height in the $y$ direction) defined as :
\begin{equation}
 h(x)=h_{\rm max} \cos ^{2}\left(\frac{\pi x}{2 l}\right), \quad-l \leq x \leq l
\end{equation}
where $h_{ \rm max}=100~$m and $l= 260~$m are respectively, the height and the half-width of the hill.
Note that, in the three cases, $x=0$ corresponds to the position of the source.
Hence the hill is shifted between case C and case B (see Fig. \ref{fig:caseB-C}).
$x=0$ corresponds to the beginning of the hill in case B and to the top of the hill in case C.
The ground impedance model used in the three cases is the variable porosity model \cite{attenborough_outdoor_2011}.
The effective flow resistivity is set to $50~$kNs.m$^{-4}$ and the effective porosity change rate to $100~$m$^{-1}$.
These values are typical for natural soil \cite{cotte_coupling_2018}.

\begin{figure}[tbp]
\centering
 \includegraphics[ width=1.00\textwidth]{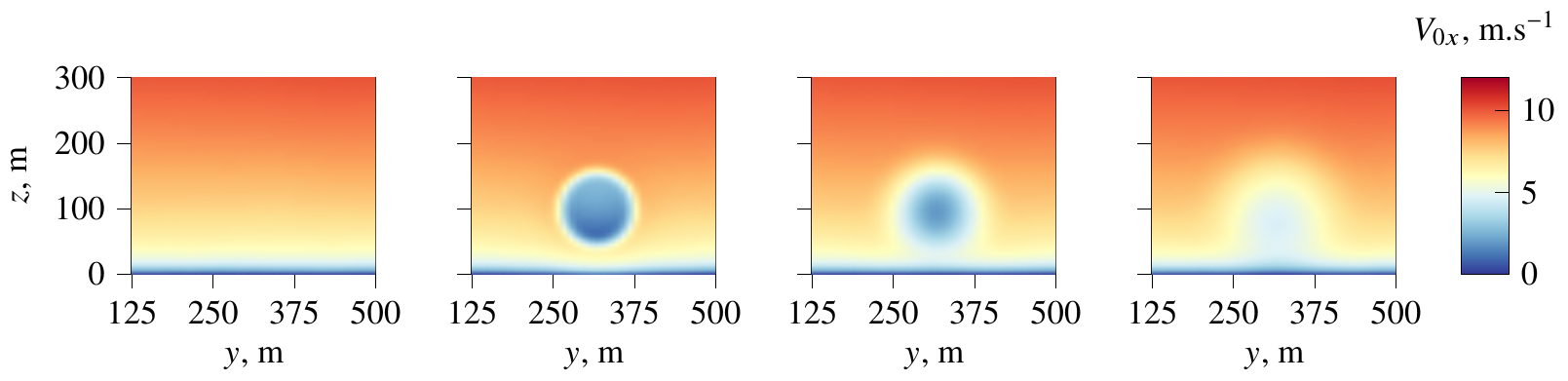}
\vspace{-0.8cm} \caption{Mean flow from LES in the $yz$ plane. From left to right at position $x=-100~$m, $x=0~$m, $x=100~$m and $x=200~$m, with $x=0~$m the turbine's location.}
\label{fig:deficit}
\end{figure}

 The side view of the axial component of the mean velocity is plotted in the plane of the wind turbine's hub in Fig. \ref{fig:caseA} for cases $\rm A_{ABL}$ and $\rm A_{WT}$.
 In the presence of the wind turbine, the velocity deficit, which arises downstream of it, takes a few hundred meters to recover.
The shape of the wake can also be observed Fig. \ref{fig:deficit} in the $yz$ plane at different axial positions.
It presents a circular zone with a few meters per second deficit just after the wind turbine, that progressively fades away with the distance from the wind turbine.

In the presence of topography the flow is first shown without the wind turbine in Fig. \ref{fig:caseB-C}.
A clear increase in speed can be observed at the top of the hill as the flow accelerates around it.
A recirculation zone is then created downstream of the hill, which leads to the reduction of the mean velocity towards negative values.
With the wind turbine in front of the hill (Fig. \ref{fig:caseB-C}) the wake of the turbine follows the shape of the hill, counterbalancing the increase of speed at the top.
Hence the hill's wake appears longer and higher.
In the case of the wind turbine on top of the hill (Fig. \ref{fig:caseB-C}), the turbine's wake is larger and merges with the hill's wake.

\section{Numerical Set-up}
\begin{figure}
 \centering
 \includegraphics{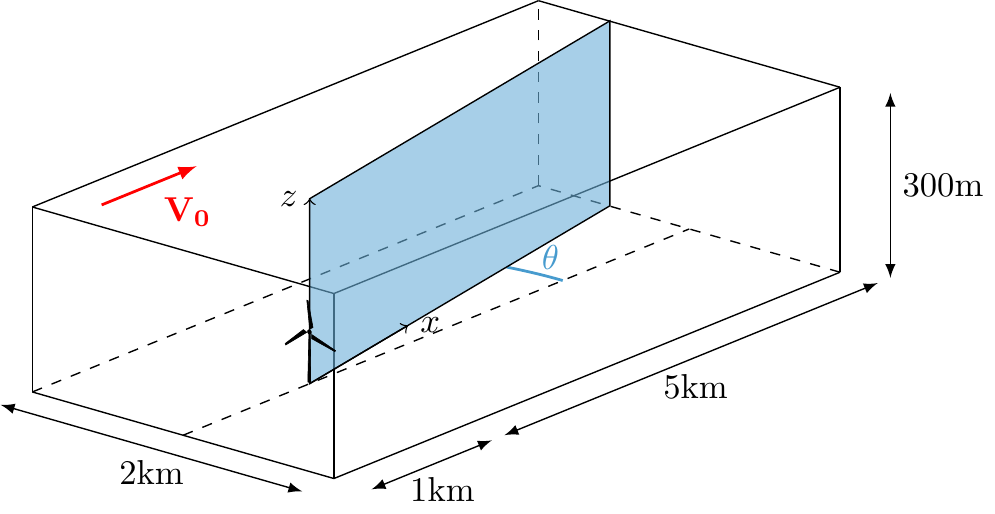}
\vspace{-0.2cm} \caption{Sketch of the computational domain. $\theta$ is the angle of propagation presented in Fig. \ref{fig:source}}
 \label{fig:domain}
\end{figure}

For each of the cases presented in Sec.~\ref{cases}, numerical simulations of the noise propagation are performed.
A $N\times$2D approach is used. This means that sound propagation is computed in vertical planes as described in \ref{Methodo}, neglecting transverse propagation effects.
The 3D rectangular domain has a size of 2~km$\times$6~km$\times300~$m as illustatred in Fig. \ref{fig:domain}.
The moving frame has a length of 300~m$\times$300~m which allows to capture most of the reflections without being too computationally expensive.
The simulations are run until the moving frame has reached the end of the domain.
For a propagation of 5~km it takes over 130000 iterations with a grid step $\Delta x = \Delta z = 0.5~$m and a CFL equal to 0.8.
Those numerical parameters produce accurate frequency results up to $1~$kHz.

The 2D simulations are performed for a set of angles of propagation $\theta$ (Fig. \ref{fig:domain}), which allows to recover a $\Delta L$ and the SPL map around the wind turbine.
Because the sound pressure levels are much higher downwind, the propagation simulations are performed up to 5~km in this direction and only up to 1km in the crosswind direction.
Hence the angular step must be smaller for the resolution downwind of the wind turbine.
For $\theta$ between 0° and 15°, a 2° angular step is used, and an angular step of 5° is used for the rest of the domain.
This is done in order to save computational resources as a 1km propagation simulation takes around $240~$CPU hours, and a full simulation (with all propagation angles) takes around 20000~CPU hours.
The results are then computed in the frequency domain and presented averaged in third octave bands.

\section{Results and Discussion }
\label{results}
In this section the results obtained from the simulations are presented and discussed.
First, the effect of the flow and topography on the acoustic propagation is investigated in each of the three cases.
Then their respective SPL are compared.
Finally the effect of the wind speed on the results is also presented.

\begin{figure}[bp]
\centering
 \includegraphics[ width=1.00\textwidth]{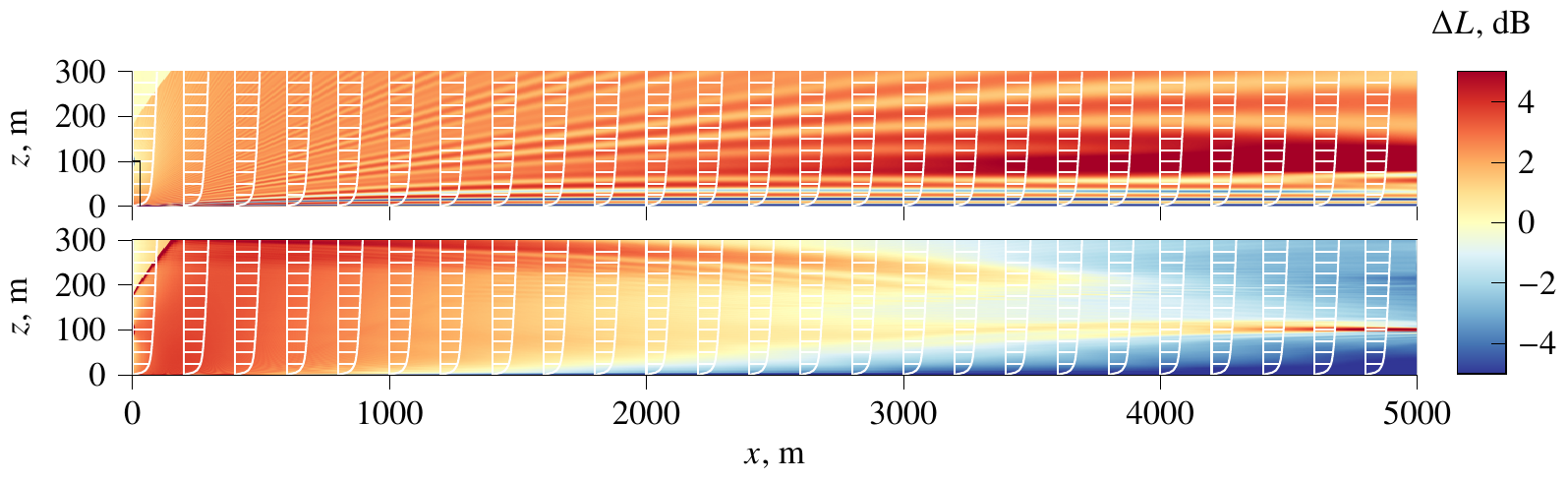}
\vspace{-0.8cm} \caption{Relative sound pressure level (color) and velocity profile (white) for case $\rm A_{ABL}$ for (top) $f_c=100~$Hz and (bottom) $f_c=1~$kHz.}
\label{fig:caseAr}
\end{figure}

\subsection{Effect of Wind Turbine Wake for Flow over Flat Terrain}
First the baseline case without a hill is presented as it serves as reference for the cases with the hill.
In Fig. \ref{fig:caseAr}, the relative sound pressure level ($\Delta L$) contour is plotted for two frequencies (100~Hz and 1~kHz) with the wind turbine sound source positioned at $x=0$ but without taking the wake effect into account.
The axial velocity profiles are also plotted in white.
Without the wind turbine in the flow and with a flat ground (case $\rm A_{ABL}$), the flow comes down to the gradient of the atmospheric boundary layer.
For both frequencies the effect of the velocity gradient is significant.
It creates a focus zone at $x=5~$km and at hub height that should be redirected toward the ground further downstream.
Because the results are presented in third octave averaged bands, the $\Delta L$ contour plotted for $f_c=1~$kHz is smoother than for $100~$Hz as it is averaged over a broader frequency band.
Still, it can be observed that the focusing is more pronounced at higher frequencies.

\begin{figure}[tbp]
\centering
 \includegraphics[ width=1.00\textwidth]{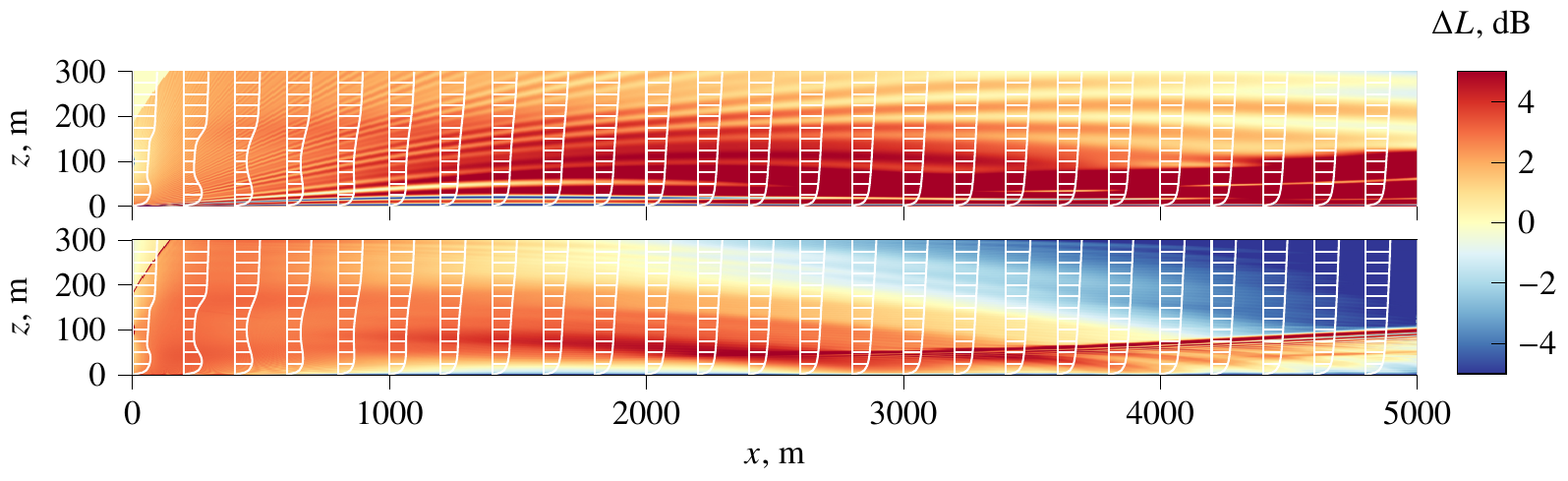}
\vspace{-0.8cm} \caption{Relative sound pressure level (color) and velocity profile (white) for case $\rm A_{WT}$ for (top) $f_c=100~$Hz and (bottom) $f_c=1~$kHz}
\label{fig:caseAb}
\end{figure}

As already studied by \citet{barlas_effects_2017} and \citet{heimann_3d-simulation_2018}, the effect of the wake on the SPL observed at the ground is known to be significant.
Figure \ref{fig:caseAb} presents equivalent colormap as Fig. \ref{fig:caseAr} but with an ABL perturbed by the wind turbine.
It is very clear, comparing both figures, that the wake has a strong influence on sound propagation.
The velocity deficit downstream of the wind turbine acts as a waveguide and leads to the focusing of acoustic waves at about 3~km from the source.
Part of this energy is then redirected towards the ground.
 With the wake taken into account, the focusing happens much closer to the source and to the ground which leads to an increase in SPL at the ground.

\begin{figure}[tbp]
\centering
 \begin{subfigure}[b]{0.49\textwidth}
 \centering
 \includegraphics[width=0.95\textwidth]{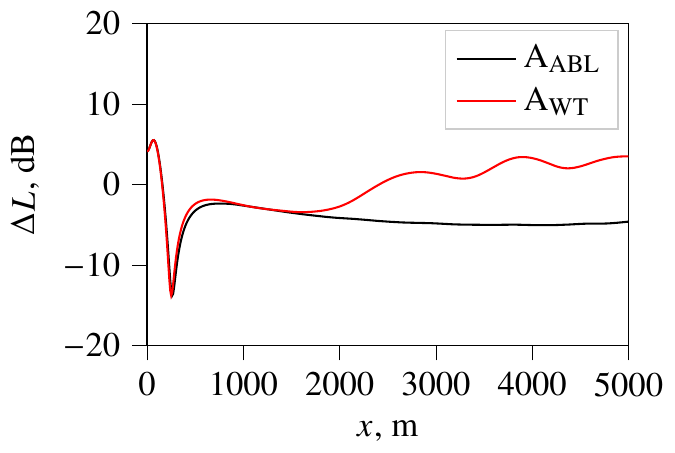}
% \vspace{0.6cm}
 \end{subfigure} 
 \begin{subfigure}[b]{0.49\textwidth}
 \centering
 \includegraphics[width=0.95\textwidth]{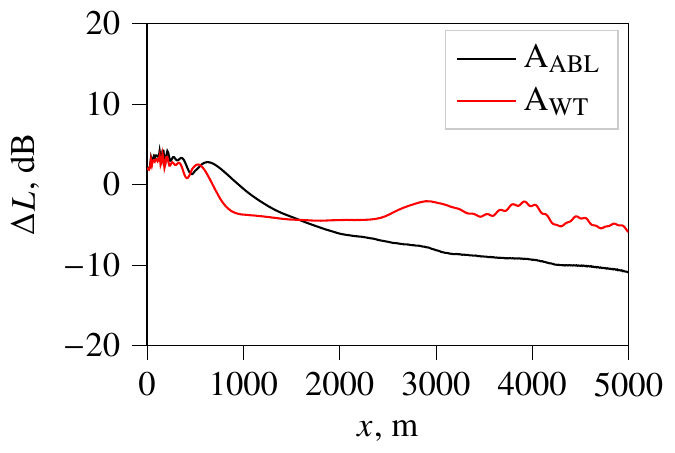}
% \vspace{0.6cm}
 \end{subfigure} 
\vspace{-0.4cm} \caption{Relative sound pressure level at 2~m height for case A with and without the turbine at (left) $f_c=100~$Hz and (right) $f_c=1~$kHz. }
\label{fig:caseArec}
\end{figure}

The relative sound pressure levels at 2~m height receivers are shown for cases $\rm A_{ABL}$ and $\rm A_{WT}$ in Fig.~\ref{fig:caseArec}.
A net increase in $\Delta L$ (over 5~dB) due to the wake can be observed at both frequencies.
A very similar behavior is observed at both frequencies with three peaks in $\Delta L$ between $x=2.5~$km and $x=5~$km.
The maxima are not aligned for all frequencies and hence will tend to average out when looking at the overall sound pressure level.

In agreement with literature \cite{barlas_effects_2017,heimann_sound_2018} our results show that wind turbine wake focuses the sound toward the ground and increase the sound pressure level in the far field.

\subsection{Effect of the Hill}
\begin{figure}[tbp]
 \centering
 \includegraphics[ width=1.00\textwidth]{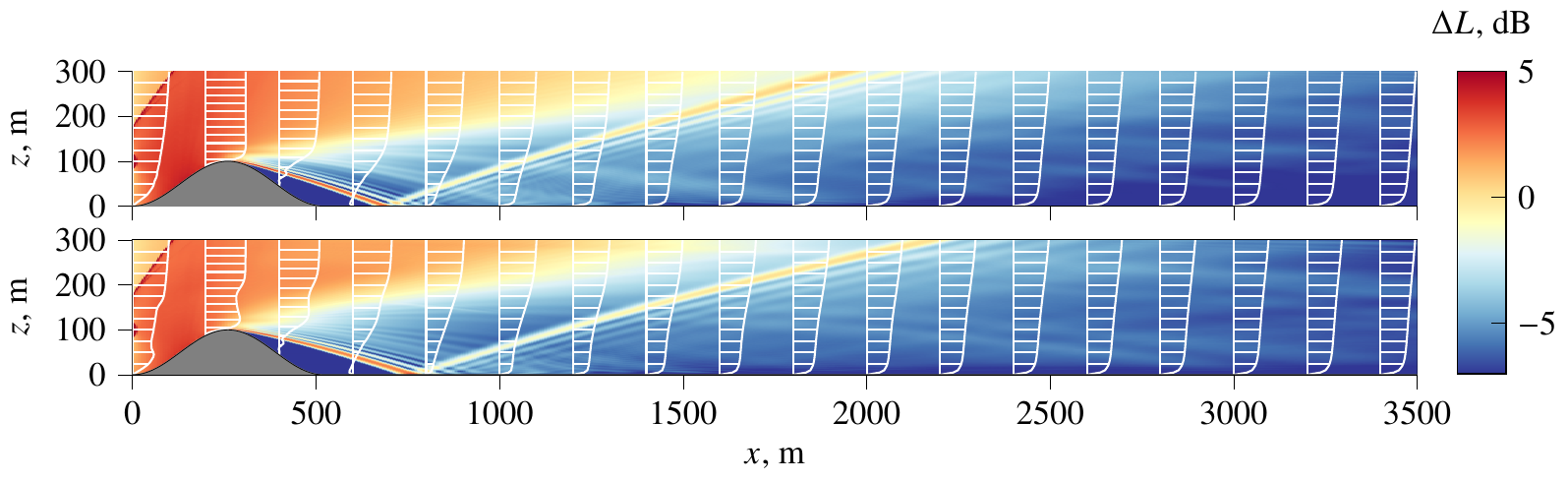}
\vspace{-0.8cm} \caption{Relative sound pressure level (contour) and velocity profile (white) for (top) case $\rm B_{ABL}$ and (bottom) $\rm B_{WT}$ for $f_c=1~$kHz.}
\label{fig:caseB}
\end{figure}

The aim of this section is to differentiate the effect on propagation brought by the ABL, the wake and the hill.
As discussed in Sec. \ref{cases}, in the case of a non flat terrain the hill will affect the propagation both directly, through reflections and diffraction of the acoustic wave by the terrain, and indirectly, by modifying the mean flow.
Hence the effect of the wake presented for the baseline case can be enhanced or balanced out by the topographical effect.
For the case with the wind turbine positioned in front of the hill, the contour of $\Delta L$ for the case with ($\rm B_{WT}$) and without ($\rm B_{ABL}$) the wake are presented in Fig. \ref{fig:caseB}.
It seems here that the hill can explain most of the visible effects and that the wake has very little influence on sound propagation.
First a shadow zone is created just behind the hill.
Second, the strong velocity gradient created just after the hill is responsible for the focusing that can be observed at 700~m at ground level (Fig. \ref{fig:caseB}).
However, even with this redirection caused by the mean flow, the shielding effect of the hill is strong.
Hence $\Delta L$ is negative after the hill, with the exception of this focusing zone.
In comparison the velocity gradient induced by the wake itself is small (see the velocity profile in Fig. \ref{fig:caseB}).
Hence the wave guide effect that was observed for case $\rm A_{WT}$ is not visible.
This indicates that the effect of the wake created by a turbine placed just upstream of a hill is overshadowed by the effect of the hill itself.
Nevertheless, a slight shift of the focused ray away from the hill can be observed in case $\rm B_{WT}$.

\begin{figure}[tbp]
 \centering
 \includegraphics[ width=1.00\textwidth]{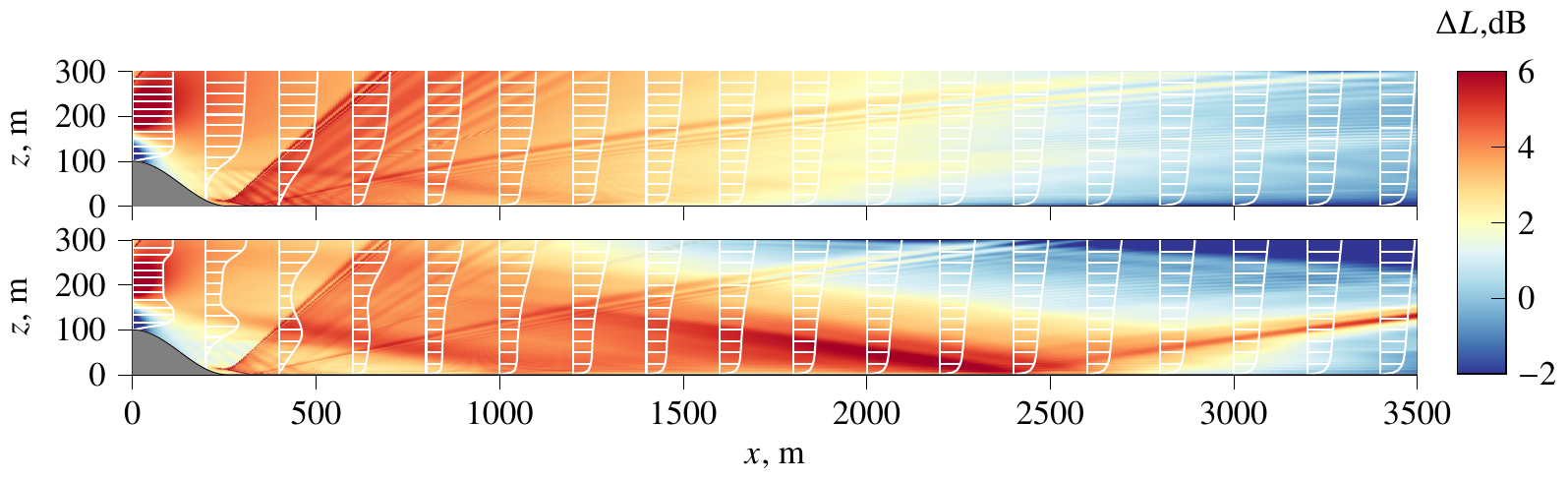}
\vspace{-0.8cm} \caption{Relative sound pressure level (contour) and velocity profile (white) for (top) case $\rm C_{ABL}$ and (bottom) $\rm C_{WT}$ for $f_c=1~$kHz }
\label{fig:caseC}
\end{figure}

However, when the wind turbine is placed on top of the hill (cases $\rm C_{ABL}$ and $\rm C_{WT}$), the wind turbine wake has a very strong effect, see Fig. \ref{fig:caseC}.
As for case B, the hill influences the propagation both through its geometry and the induced mean flow.
Here, the geometry of the hill creates a cusp caustic just above the bottom of the hill (at 260~m on Fig. \ref{fig:caseC}).
The caustic then separates into two branches, one going directly up and one reflecting towards the ground at 330~m.
This is only an effect of the terrain and can be observed for both cases.
The second effect is similar to what was observed in case B.
The velocity deficit behind the hill creates a focusing zone at the ground at $x=1300~$m.
The effect is less visible than for case B because the direct influence of the source is stronger.
Next, looking at $\Delta L$ obtained with the wake taken into account (Fig. \ref{fig:caseC}), some differences can be noticed.
Firstly the focusing induced by the hill's wake seems much stronger and appears closer to the source, from $x=1200~$m to $x= 1000~$m.
This effect seems to be a combination of the geometry and the merging of the wakes from the turbine and the hill.
Secondly, another focusing appears, which is redirected to the ground at 2400~m.
This is a direct consequence of the wake, which is acting as a wave guide in the same way it was in the flat case.
As a consequence the high $\Delta L$ region spreads over a larger distance at ground-level in case $\rm C_{WT}$ than $\rm C_{ABL}$.

\begin{figure}[tbp]
\centering
 \begin{subfigure}[b]{0.49\textwidth}
 \centering
 \includegraphics[width=0.95\textwidth]{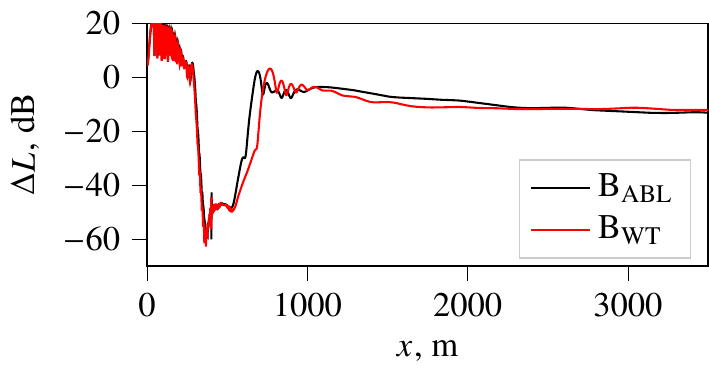}
% \vspace{0.6cm}
 \end{subfigure} 
 \begin{subfigure}[b]{0.49\textwidth}
 \centering
 \includegraphics[width=0.95\textwidth]{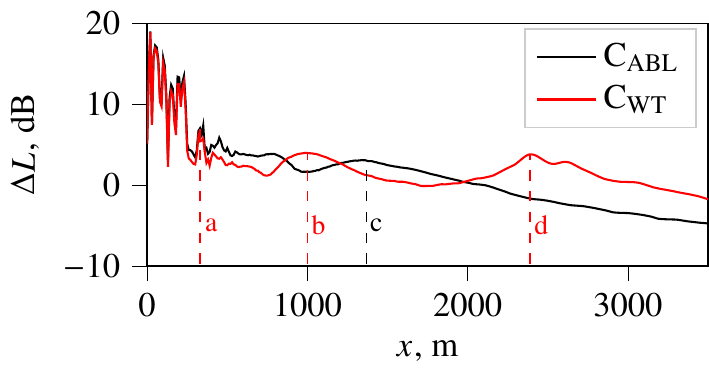}
% \vspace{0.6cm}
 \end{subfigure} 
\vspace{-0.4cm} \caption{Relative sound pressure level at 2m height for (left) case B and (right) C at $f_c = 1~$kHz, with and without the wind turbine.}
\label{fig:topoLine}
\end{figure}

The relative sound pressure levels for a line of receivers at 2m height are plotted in Fig. \ref{fig:topoLine} for case A and B.
It is really clear for case B that the wake has very little influence on the SPL at the ground.
The dip between 500~m and 700~m corresponds to the shadow zone observed in Fig. \ref{fig:caseB} just behind the hill.
The shadow zone is a bit larger in the presence of the wake as the focusing seems to appear a few meters away from the hill.
It was already visible on Fig. \ref{fig:caseB} as the focused ray arrives before 700~m for case $\rm B_{ABL}$ but around 800~m for $\rm B_{WT}$
This is explained by the fact that the velocity gradient behind the hill is softened by the wake and hence the redirection towards the ground occurs a bit further.
In the far field ($x > 1~$km) no differences are noticed.

For case C different arrivals at the ground are identifiable, corresponding to the previously described effects.
The peak at $a$ corresponds to the caustic redirected towards the ground.
Peaks $b$ and $c$ correspond to the focusing induced by the hill's wake.
In case $\rm C_{WT}$ the focusing appears 200~m closer to the source than for case $\rm C_{ABL}$.
Finally, the strongest sound focusing appears at $d$, which is the one directly induced by the turbine's wake, with a 5~dB increase between cases $\rm C_{ABL}$ and $\rm C_{WT}$.
The peaks $c$ and $b$ are of comparable amplitudes and the major effect of the wake seems to be the 300~m shifting towards the source.

\begin{figure}[tbp]
 \centering
 \includegraphics[width=0.70\textwidth]{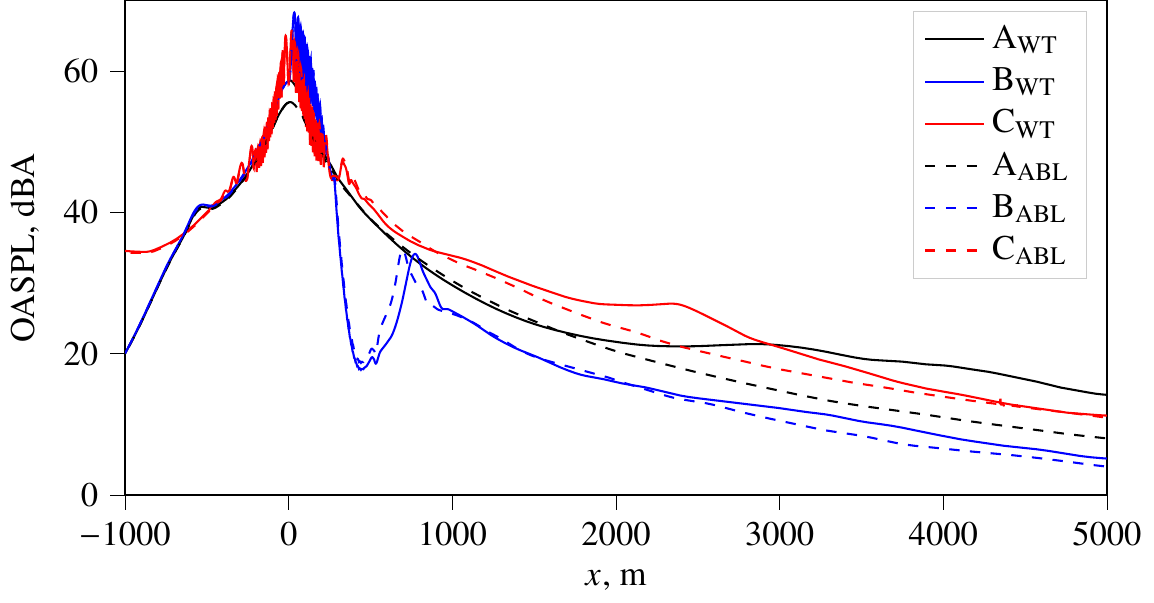}
\vspace{-0.4cm} \caption{Comparison between the case A , B and C. Results are plotted for cases with (solid lines) and without (dashed lines) the turbine}
 \label{fig:comparAllLine}
\end{figure}

\subsection{Comparison of overall SPL}

The comparison between the overall sound pressure level (OASPL) in dBA for the three cases studied is presented in Fig. \ref{fig:comparAllLine}.
First the observation made for the relative SPL for frequency bands are in accordance with overall SPL results.
For case A the increase in sound power level in presence of the wake is very visible after 2~km.
For case B, the shadow zone widening between 500~m and 800~m in the presence of the wake is coherent with previous observation.
Finally in case C, the three focusing zones (at $x=200~$m, $x=1100~$m and $x=2500~$m) described in Fig. \ref{fig:topoLine} are still distinct even if the second one is a lot smoother.

The first observation is that the overall SPL, in the downwind direction ($x>0$), are always between 5~dBA and 10~dBA smaller for case B than for the two other cases.
However, the overall SPL at the focus zone ($x=700$~m) in case B is of comparable amplitude with the SPL for the other cases at this position.
Although the hill shields the noise, levels remain significant at ground level just after the shadow zone.
In the upwind direction ($x<0$) case A and case B are equivalent as there is no topography or significant changes in the mean flow.
Without taking the wake into consideration the overall SPL is higher in case $\rm C_{ABL}$ with a consistent difference of 3~dBA between the SPL computed for case $\rm A_{ABL}$.
On the contrary taking the wake into consideration shows that the overall SPL, compared to $\rm C_{ABL}$, is higher closer to the source in case $\rm C_{WT}$ (with a peak at 2.5~km) but remains the same in the far field.
For case $\rm A_{WT}$ the overall SPL increases from 3~km to 5~km.
Hence, the overall SPL in case $\rm A_{WT}$ becomes greater than the level for case $\rm C_{WT}$ after 3~km.
This can be explained as more energy is focused close to the hill in case $\rm C_{WT}$ which leads to reduced sound pressure level in the far field, while in case $\rm A_{WT}$, focusing of acoustic waves occurs at a larger distance which leads to increased sound pressure level after 3~km.
This shows that the sound will propagate at greater distances in the downwind direction for a flat terrain configuration than for a wind turbine on top of a hill.

\begin{figure}[tbp]
\centering
 \includegraphics[ width=1.00\textwidth]{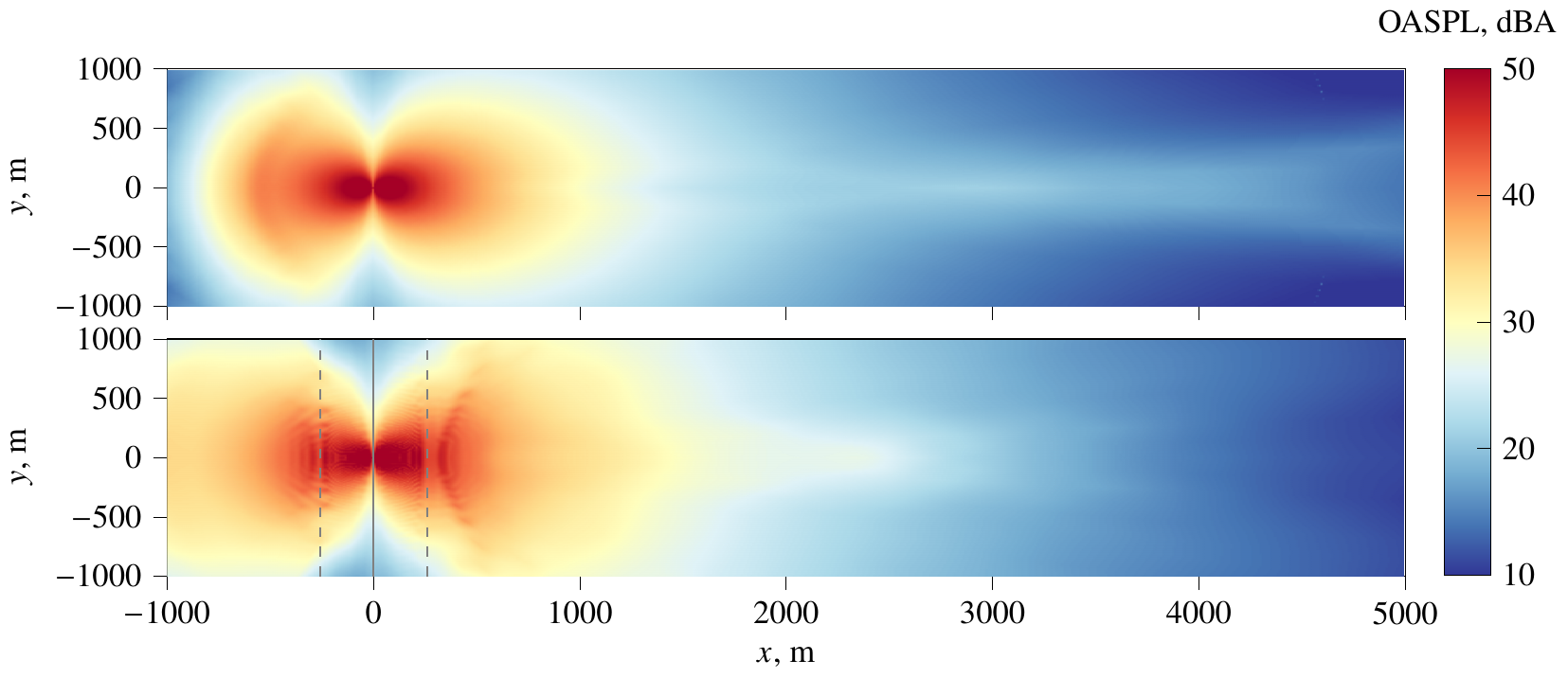}
\vspace{-0.8cm} \caption{Overall sound pressure level at 2~m height for (top) case $\rm A_{WT}$ and (bottom) case $\rm C_{WT}$. The top of the hill is represented in solid line and the bottom in dashed lines.}
\label{fig:top}
\end{figure}

An interpolated top view of the overall SPL for case $\rm A_{WT}$ and $\rm C_{WT}$ at 2m height is presented in Fig. \ref{fig:top}.
It confirms that, in both cases, the high SPL zones are present downwind (where the sound waves propagate through the wake of the turbine) with a width of 200~m.
As previously described, more energy is directed before 3~km in the case with the hill while the focusing zone is present at much greater distances for case A.
Finally it can be observed upwind of the turbine that the overall SPL are higher for case C.
This is an effect of the elevation of the source, that will tend to move back the shadow zone created by mean flow.
There is no difference upwind between the cases with or without the wake as the sound does not propagate through it.

\begin{figure}[tbp]
\centering
 \includegraphics[ width=1.00\textwidth]{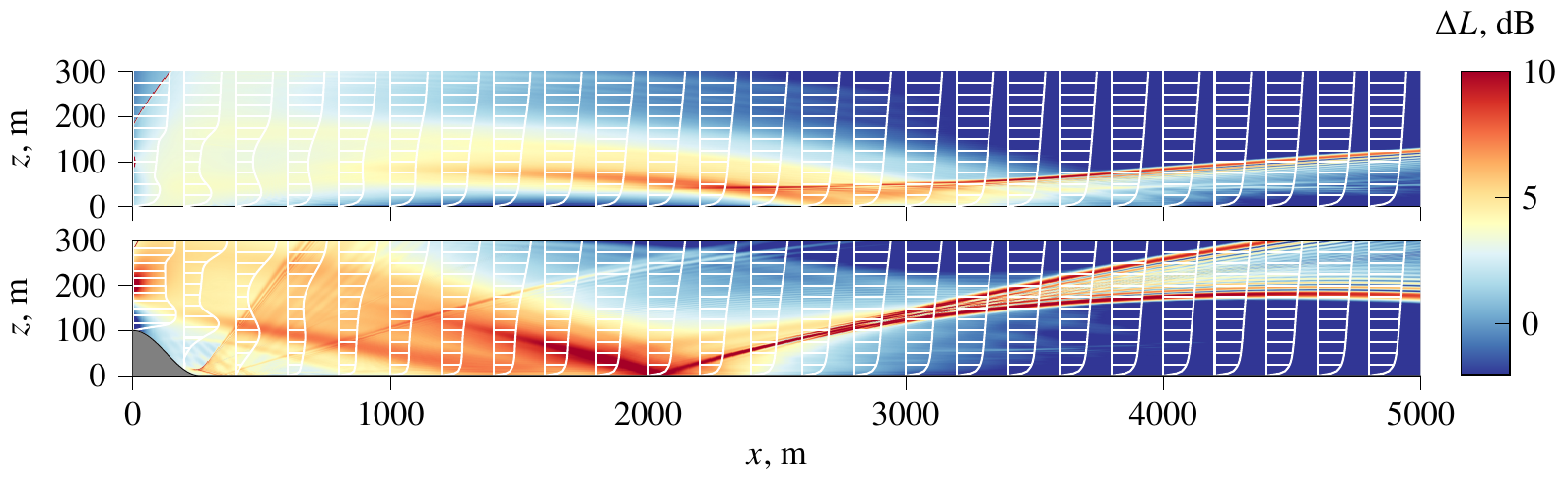}
\vspace{-0.8cm} \caption{Contour of $\Delta L$ at 1~kHz for $V_{0x}(100~$m$) = 12~$m.s$^{-1}$ for case $\rm A_{WT}$ (top) and $\rm C_{WT}$ (bottom). The wind speed profile is superimposed in white.}
\label{fig:windSpeedcontour}
\end{figure}

\subsection{Influence of Velocity}
In this last section the effect of the velocity on the propagation is briefly examined.
The previous results were obtained for a wind profile characterized by a wind speed at 100~m (hub height) equal to $V_{\rm hub} = 8~$m.s$^{-1}$.
In general a wind turbine operates at wind speeds between 6~m.s$^{-1}$ and 14~m.s$^{-1}$.
These more extreme cases should also be investigated as propagation effects are expected to be more important.
Two additional simulations are run for case $\rm A_{WT}$ and $\rm C_{WT}$ with $V_{\rm hub}$ equal to 10~m.s$^{-1}$ and 12~m.s$^{-1}$.
These modifications are made from the available LES data by scaling the flow fields with $u^*$.
This rough approximation allows to have a first idea of the influence of the velocity on the propagation without consuming more computational resources.
Note that, for comparison, the sound power level is kept the same, while it is expected that the wind turbine actually emits more noise with the increase of the wind speed at hub height.

In the case of $V_{\rm hub}=12~$m.s$^{-1}$, the $\Delta L$ contour plot at 1~kHz for cases $\rm A_{WT}$ and $\rm C_{WT}$ are presented Fig.~\ref{fig:windSpeedcontour}.
The effects of the mean flow are clearly more intense compared to the wind speed of $V_{\rm hub}=8~$m.s$^{-1}$ (Fig. \ref{fig:caseAb} and Fig. \ref{fig:caseC}).
Note that the colormap range is also different between plots.
The general dynamics are similar but the focusing in case A is much stronger and, in case C, the three different focus zones are now more distinguishable.
These three focusing zones seem to be related to the high velocity gradient zones (I) around the hilltop, and at the (II) bottom and (III) top of the wake.
These focusing are redirected at some point towards the ground by the general positive velocity gradient of the ABL.
The separation of the strongest ray after bouncing of the ground is due to the velocity gradient that folds the wavefront and creates an other caustic.
These phenomena could also be observed for $V_{\rm hub}=10~$m.s$^{-1}$ but the effect is less pronounced.

\begin{figure}[tbp]
\centering
 \begin{subfigure}[b]{0.49\textwidth}
 \centering
 \includegraphics[width=0.95\textwidth]{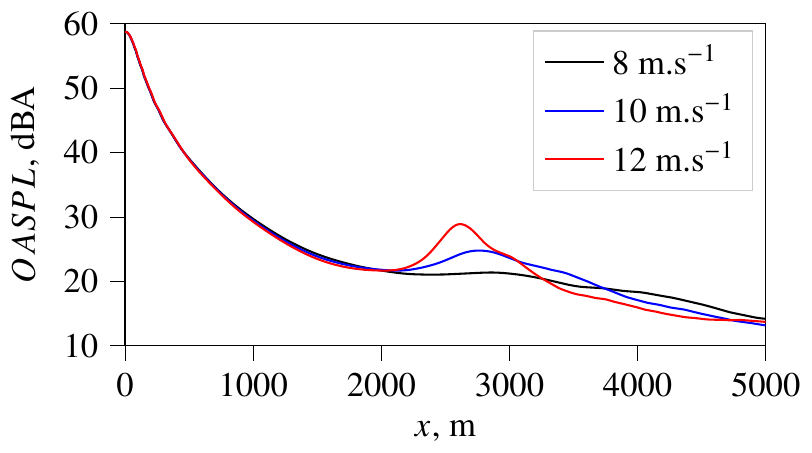}
% \vspace{0.6cm}
 \end{subfigure} 
 \begin{subfigure}[b]{0.49\textwidth}
 \centering
 \includegraphics[width=0.95\textwidth]{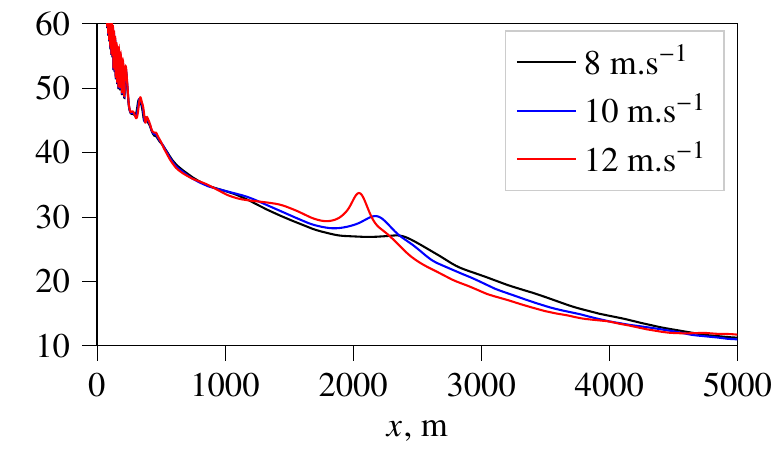}
% \vspace{0.6cm}
 \end{subfigure} 
\vspace{-0.4cm} \caption{Overall SPL for cases A (left) and C (right) for different wind speeds at hub height as indicated in the legend.}
\label{fig:windSpeed}
\end{figure}

The overall sound pressure levels for receivers at 2 m height above the ground are presented for the three wind speeds in Fig. \ref{fig:windSpeed}.
The same behavior can be observed for case A and case C.
By increasing the wind speed the focusing phenomenon tends to intensify: the peak is stronger and more localized.
There is a 10~dBa increase for case A between $V_{ \rm hub} = 8~$m.s$^{-1}$ and $V_{ \rm hub} = 12~$m.s$^{-1}$.
This effect is smaller in case C with a 7~dBA increase between $V_{ \rm hub} = 8~$m.s$^{-1}$ and $V_{ \rm hub} = 12~$m.s$^{-1}$.
As the effective sound speed increases the focusing at the ground also appears closer to the source.
For case C, the maximum shifts 500~m closer to the source between the case at 8~m.s$^{-1}$ and 12~m.s$^{-1}$.
As the wind speed increases, the overall SPL after the focused zone decreases.
This can be explained as the focusing becomes stronger and more localized there is less energy that is redirected at long distance.

\section{Conclusion}

This study demonstrates that terrain topography can significantly affect wind turbine sound propagation.
Using linearized Euler equations, solved in a moving frame following the wavefront, we find that the velocity gradient in wind turbine wakes leads to a sound focalization toward the ground.
For turbines on a hilltop, the sound pressure level due to the turbine shows more substantial peaks closer to the turbine than for a reference flat terrain case.
However, when the turbine is placed before the hill, the effect of the wind turbine wake is limited as sound propagation is mainly determined by the geometry of the terrain and the flow around the hill.
Furthermore, we find that the sound focusing effect becomes more pronounced and closer to the turbine with increasing wind speed.
Given the importance of understanding the impact of wind turbine noise propagation in different environments, including social impacts, it is crucial to understand the interactions between wind turbines and varying terrains.
A future research area of interest is exploring the effect of an extended sound source instead of a point source, as this would allow to study the impact of topography on amplitude modulation.

\section*{Acknowledgements}
We thank Luoqin Liu for providing us access to the LES data of ref. \cite{liu_effects_2020}. This work was performed within the framework of the LABEX CeLyA (ANR-10-LABX-0060) of Universit\'e de Lyon, within the program ``Investissements d’Avenir" (ANR-16-IDEX-0005) operated by the French National Research Agency (ANR). It was granted access to the HPC resources of PMCS2I (P\^ole de Mod\'elisation et de Calcul en Sciences de l'Ing\'enieur et de l'Information) of Ecole Centrale de Lyon, PSMN (P\^ole Scientifique de Mod\'elisation Num\'erique) of ENS de Lyon and P2CHPD (P\^ole de Calcul Hautes Performances D\'edi\'es) of Universit\'e Lyon I, members of FLMSN (F\'ed\'eration Lyonnaise de Mod\'elisation et Sciences Num\'eriques), partner of EQUIPEX EQUIP@MESO. This work is part of the Shell-NWO/FOM-initiative Computational sciences for energy research of Shell and Chemical Sciences, Earth and Live Sciences, Physical Sciences, FOM and STW, and an STW VIDI grant (No. 14868).

%\bibliography{sample}

\end{document}